\documentclass[aps,pra,twocolumn,groupedaddress,showpacs]{revtex4-1}
\usepackage{graphicx}
\usepackage{leftidx}
\usepackage{epstopdf}

\begin{document}


\title{Goos-H\"{a}nchen and Imbert-Fedorov Shifts of Vortex Beams at Air-Left-Handed Material Interfaces}


\author{Zhicheng Xiao}
\author{Hailu Luo}
\email[]{hailuluo@hnu.edu.cn}
\author{Shuangchun Wen}
\email[]{scwen@hnu.edu.cn}
\affiliation{Key Laboratory for Micro/Nano Optoelectronic Devices of
Ministry of Education\\School of Information Science and
Engineering, Hunan University, Changsha 410082, China}


\date{\today}

\begin{abstract}
In this paper, we present a systematic study of beam shifts and angular momenta of paraxial vortex beams at air-left-handed material (LHM) interfaces. It is shown that, compared to their counterparts at air-right-handed material (RHM) interfaces, the spatial Goos-H\"{a}nchen (GH) and Imbert-Fedorov (IF) shifts remains the same, while the angular GH and IF shifts are reversed, at air-LHM interfaces. The spatial and angular shifts of paraxial vortex beams have their respective origins in transverse angular momenta and transverse linear momenta. The spatial GH and IF shifts remain unreversed as a result of the both reversions of transverse angular momenta and z-component linear momentum, while the angular GH and IF shifts are reversed due to that the z-component linear momentum is reversed and the transverse linear momenta are unreversed, at air-LHM interfaces. In addition, we perform a quantitative analysis on spin-orbit angular momentum conversion and orbit-orbit angular momentum conversion, which further helps us understand the essence of vortex beam shifts at air-LHM interfaces and their fundamental distinctions with those at air-RHM interfaces.
\end{abstract}


\maketitle

\section{Introduction}{\label{introduction}}
The reflection and transmission of light at air-medium interface has been a topic of ongoing investigation and contention. Initially, the amplitudes and directions of reflected and transmitted light were determined by Fresnel formulas and Snell's law. Deviations were later found in total internal reflection (TIR), however. They are Goos-H\"{a}nchen (GH) shift~\cite{Goos-hanchen1947,Artmann1948,Lai1986,Lima2011} and Imbert-Fedorov (IF) shift~\cite{Fedorov1955,Schilling1965,Imbert1972,Player1987,Menard2010,Dartora2011}, which are parallel and perpendicular to the incident plane, respectively. Artmann formula~\cite{Artmann1948} for GH shift and Schilling formula~\cite{Schilling1965} for IF shift in TIR are widely accepted. The exact formulas for IF shifts in partial reflection and transmission, however, have divided many physicists~\cite{Player1987,Pillon2004,Onoda2004,Bliokh2006,Bliokh2007}. Owing to Hosten's precise measurement, the debate was settled down~\cite{Hosten2008}. Apart from the constant spatial GH and IF shifts, the light beam also experiences angular shifts~\cite{Chan1985,Merano2009}, which increase proportionally
with propagation distance $z$. In general, the spatial GH and IF shifts stem from spin-orbit interaction at the air-medium interface~\cite{Bliokh2009,Fedoseyev2011}. To satisfy the $z$-component angular momentum conservation law, the reflected and transmitted beams must possess extrinsic transverse orbital angular momenta, which results in spatial GH and IF shifts. The angular shift is essentially a diffractive correction on light beams and governed by the linear momentum conservation law~\cite{Bliokh2009,Fedoseyev2009}.

The emergence of left-handed materials (LHM) has brought about great opportunities and sophisticated pathways to manipulate light~\cite{Pendry1996,Shelby2003,Lezec2007,Liu2009}. As a branch of metamaterials, LHM usually refers to those materials whose permittivity and permeability are negative simultaneously. It has shown very unique properties, like negative refraction~\cite{Lezec2007}, inverse Doppler effect~\cite{Seddon2003,Chen2011}, unreversed rotational Doppler effect~\cite{Luo2008} and inverse Cherenkov radiation~\cite{Duan2008}. Apart from these properties, beam shifts in LHM have been demonstrated to be significantly different from right-handed materials (RHM)~\cite{Berman2002,Shadrivov2003,Menzel2008,Krowne2009,Luo2009}. Theoretical papers~\cite{Berman2002,Shadrivov2003} indicate that light beam experiences negative GH shift at air-LHM interface and this shift can be amplified in layered structure. The IF shift in partial reflection, however, remains unreversed owing to unreversed spin angular momentum in LHM~\cite{Luo2009}. As for vortex beam, it carries intrinsic orbital angular momentum. Therefore, orbit-orbit conversion is inevitable in reflection and transmission. Will the spatial GH and IF shifts remain unreversed in the presence of orbital angular momentum? How about the angular shifts? What does the physical picture of spin-orbit and orbit-orbit conversions look like? Clarifying these problems is not only conducive to understanding the detailed behaviors of vortex beams at air-LHM interface but also sheds a new light on the essence of angular momentum and linear momentum of vortex beam.

In this paper, we endeavor to offer concrete expressions of spatial and angular shifts of vortex beam at air-LHM interface and elaborate on their relations with the linear and angular momenta. We also contrast these results with air-RHM interface. Although some of the results are similar to our previous paper~\cite{Luo2009}, we conduct considerably rigorous analyses on the topics and thorough explanations on the results. The rest of the paper is arranged as follows. In Sec.~\ref{electric fields}, we adopt angular spectrum method to derive the electric fields of reflected and transmitted vortex beams at air-LHM interface. Except for some special cases, for instance, in the vicinity of Critical angle of TIR and Brewster angle, the expressions of electric fields generally hold true. The longitudinal fields are included as well, which take on polarization-sensitive vortex structure~\cite{Bekshaev2012}, providing a new perspective on the mechanism of IF shifts other than spin-orbit conversion. In Sec.~\ref{shifts}, we adopt the operator method to calculate the spatial and angular shifts, which is significantly different from conventional calculation method~\cite{Bliokh2007,Aiello2008,Luo2009,Bekshaev2012} and save us considerable time and efforts. The GH and IF shifts in TIR and partial transmission are demonstrated and contrasted with air-RHM interface. The impact of incident angle on beam shifts is analyzed as well. In Sec.~\ref{momenta}, we calculate the linear and angular momenta of incident, reflected, and transmitted beams, respectively. The origins of spatial and angular shifts are clarified. The quantitative analysis on spin-orbit conversion and orbit-orbit conversion is also demonstrated.

\section{Electric Fields of Reflected and Transmitted Vortex Beams}\label{electric fields}
In this study, we adopt the angular spectrum method to establish a model for the reflected and transmitted vortex beams at air-LHM interface. The main procedures of this method are as follows. We first decompose the vortex beams into plane waves with finite spectral width. Then, we analyze the incident angle and Fresnel
coefficients of each plane wave component separately. In this case, the incident angle and Fresnel coefficients of each plane wave slightly differ from that of the main Fourier component (also known as central wave component). Therefore, we expand the Fresnel coefficients in Taylor series around the central incident angle and make approximation to the first order. Afterwards, we transform the electric fields from momentum space to position space.

\begin{figure}
\includegraphics[width=8cm]{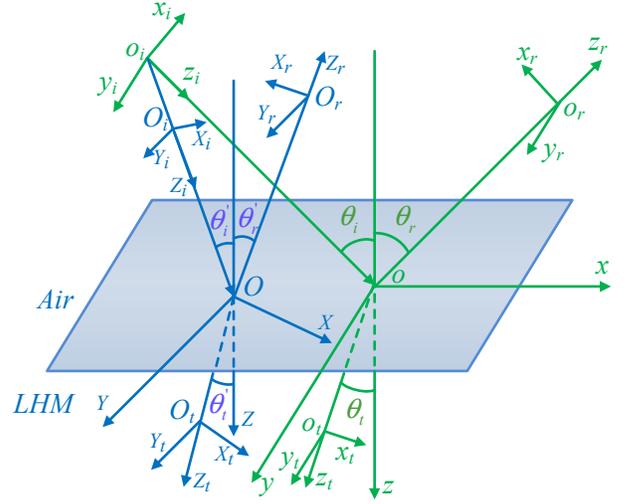}%
\caption{(Color online) Geometry of reflection and transmission. The incident plane
of the main Fourier component is $xoz$ (marked with green
lines). For an arbitrary plane wave, the incident plane is $XOZ$
(marked with blue lines).}\label{Geometry}
\end{figure}

The geometry of reflection and transmission is demonstrated in Fig.~\ref{Geometry}. The incident vortex beam propagates along $z_i$ axis and impinges on the air-LHM interface. The incident plane of the main Fourier component is $xoz$. The reflected wave and transmitted wave of the main Fourier component travel along the $z_r$ axis and $z_t$ axis, respectively. The incident angle, reflection angle, and transmission angle of the main Fourier component are $\theta_i$, $\theta_r$, and $\theta_t$, respectively. The relations among the coordinates $o-xyz$, $o_i-x_iy_iz_i$, $o_r-x_ry_rz_r$, and $o_t-x_ty_tz_t$ are determined by the Snell's Law:
\begin{equation}
\left[\matrix{\hat{\mathbf{x}}_\tau\cr\hat{\mathbf{y}}_\tau\cr
\hat{\mathbf{z}}_\tau}\right]=\left[\matrix{\cos\vartheta_\tau&0&-\sin\vartheta_\tau\cr
0&1&0\cr\sin\vartheta_\tau&0&\cos\vartheta_\tau}\right]\left[\matrix{\hat{\mathbf{x}}
\cr\hat{\mathbf{y}}\cr\hat{\mathbf{z}}}\right],\label{Snell's Law}
\end{equation}
where $\tau=i,~r$, or $t$,
$\left[\matrix{\hat{\mathbf{x}}&\hat{\mathbf{y}}&\hat{\mathbf{z}}}\right]$
and
$\left[\matrix{\hat{\mathbf{x}}_{\tau}&\hat{\mathbf{y}}_{\tau}&\hat{\mathbf{z}}_{\tau}}\right]$
are the unit basis vectors of the coordinates $o-xyz$ and
$o_\tau-x_\tau y_\tau z_\tau$, respectively,
$\left[\matrix{\vartheta_i&\vartheta_r&\vartheta_t}\right]=\left[\matrix{\theta_i&\pi-\theta_i&-\theta_t}\right]$,
$\theta_i=\theta_r$ and $\sin\theta_t=\sin\theta_i/|n|$. Constant
$n$ is the refractive index of the LHM.
For an arbitrary plane wave, we assume that the incident plane is $XOZ$. The incident, reflected, and transmitted plane wave travel along the $Z_i$, $Z_r$, and $Z_t$ axis, respectively. The incident angle, reflection angle, and transmission angle of an arbitrary Fourier component are $\theta_i^{'}$, $\theta_r^{'}$, and
$\theta_t^{'}$, respectively. The relations among coordinates $O-XYZ$, $O_i-X_iY_iZ_i$, $O_r-X_rY_rZ_r$, and $O_t-X_tY_tZ_t$ are easily acquired by revising
$\left[\matrix{\vartheta_i&\vartheta_r&\vartheta_t}\right]$ in Eq.~(\ref{Snell's Law}) as $\left[\matrix{\theta_i^{'}&\pi-\theta_i^{'}&-\theta_t^{'}}\right]$.
From Fig.~\ref{Geometry}, we derive the incident angle of an arbitrary plane wave, $\theta_i^{'}=\arccos\left(-k_{x_i}\sin\theta_i/k+k_{z_i}\cos\theta_i/k\right)$, where $k$ is the wave number in vacuum, $k_{x_i}$, $k_{y_i}$, $k_{z_i}$ are the components of wave vector along the $x_i$, $y_i$, $z_i$ axis, respectively. We can expand the incident angle $\theta_i^{'}$ around the central incident angle $\theta_i$ in series of $k_{x_i}/k$ and $k_{y_i}/k$. Therefore, $\theta_i^{'}\approx\theta_i+k_{x_i}/k$, $\theta_t^{'}\approx\theta_t+k_{x_i}/(|n|\eta k)$, where $\eta=\cos\theta_t/\cos\theta_i$.

After introducing the geometry of reflection and transmission, we start analyzing the angular spectrum of reflected and transmitted beams by using transformation matrix. Note that the incident, reflected, and transmitted vortex beams are presented in local coordinate systems $o_{\tau}-x_{\tau}y_{\tau}z_{\tau}$. The angular spectrum of incident beam is:
\begin{eqnarray}
\tilde{\mathbf{E}}_i&=&\left[\alpha\hat{\mathbf{x}}_i+\beta\hat{\mathbf{y}}_i-\frac{1}{k}\left(\alpha
k_{x_i}+\beta
k_{y_i}\right)\hat{\mathbf{z}}_i\right]\tilde{u}_i,\nonumber\\
\tilde{u}_i&=&\frac{C_l w_0}{2}\left[w_0 \left(-i
k_{x_i}+\text{sgn}[l]k_{y_i}\right)/\sqrt{2}\right]^{|l|}\nonumber\\
&&\times\exp\left[-w_0^2\left(k_{x_i}^2+k_{y_i}^2\right)/4\right],\label{incidentbeam}
\end{eqnarray}
where $\tilde{u}_i$ is the angular spectrum of vortex beams, $l$ is the vortex charge, $C_l=\sqrt{2/(\pi|l|!)}$ is the normalization constant, $w_0$ is the width of beam waist, $k_{x_i}$ and $k_{y_i}$ are wave vector components along the $x_{i}$ and $y_{i}$ axis, $k$ is wave number in vacuum, $\text{sgn}[l]$ is the sign function, $\alpha$ and $\beta$ are Jones vectors, $|\alpha|^2+|\beta|^2=1$. There are two parameters that characterize the polarization state of paraxial beams: $\sigma=2 ~\text{Im}\left[\alpha^{*}\beta\right], \chi=2 ~\text{Re}\left[\alpha^{*}\beta\right]$. $\sigma$ is the degree of circular polarization. A value of $\sigma=+1$ corresponds to left circularly polarized light beam, whereas a parameter of $\sigma=-1$ stands for right circularly polarized light beam. $\sigma=0$ represents linear polarization and values between 0 and 1 should correspond to elliptically polarized states. $\chi$ is the degree of linear polarization. It is generally recognized that the circularly polarized vortex beam has intrinsic angular momentum $(l+\sigma)\hbar$ per photon~\cite{Allen1992, Neil2002}.

We first write the electric field in coordinate $O_i-X_iY_iZ_i$. This target would be achieved in three steps. First, we write the electric field in $o-xyz$ by using the matrix in Eq.~(\ref{Snell's Law}). Second, we transform the electric field from $o-xyz$ to $O-XYZ$. This step is accomplished by using the relations:
$\hat{\mathbf{X}}=(\hat{\mathbf{z}}\times\hat{\mathbf{k}}_i)\times\hat{\mathbf{z}}/|(\hat{\mathbf{z}}\times\hat{\mathbf{k}}_i)\times\hat{\mathbf{z}}|$,
$\hat{\mathbf{Y}}=\hat{\mathbf{z}}\times\hat{\mathbf{k}}_i/|\hat{\mathbf{z}}\times\hat{\mathbf{k}}_i|$, $\hat{\mathbf{Z}}=\hat{\mathbf{z}}$, where $\hat{\mathbf{k}}_i=\left(k_{x_i}\hat{\mathbf{x}}_i+k_{y_i}\hat{\mathbf{y}}_i+k_{z_i}\hat{\mathbf{z}}_i\right)/k$. Therefore, this transformation matrix is given below:
\begin{equation}
\left[\matrix{\hat{\mathbf{X}}\cr\hat{\mathbf{Y}}\cr
\hat{\mathbf{Z}}}\right]=\left[\matrix{\frac{k_{x_i}\cos\theta_i+k_{z_i}\sin\theta_i}{k\sin\theta_i^{'}}&\frac{k_{y_i}}{k\sin\theta_i^{'}}&0\cr
-\frac{k_{y_i}}{k\sin\theta_i^{'}}&\frac{k_{x_i}\cos\theta_i+k_{z_i}\sin\theta_i}{k\sin\theta_i^{'}}&0\cr
0&0&1}\right]\left[\matrix{\hat{\mathbf{x}}
\cr\hat{\mathbf{y}}\cr\hat{\mathbf{z}}}\right].\label{matrix1}
\end{equation}
Third, we transform the electric field from $O-XYZ$ to $O_i-X_iY_iZ_i$ using revised version of Eq.~(\ref{Snell's Law}). The transformation matrix of these three steps is written as
\begin{equation}
\left[\matrix{\hat{\mathbf{X}_i}\cr\hat{\mathbf{Y}_i}\cr
\hat{\mathbf{Z}_i}}\right]=\left[\matrix{1&\frac{k_{y_i}\cot\theta_i}{k}&-\frac{k_{x_i}}{k}\cr
-\frac{k_{y_i}\cot\theta_i}{k}&1&-\frac{k_{y_i}}{k}\cr
\frac{k_{x_i}}{k}&\frac{k_{y_i}}{k}&1}\right]\left[\matrix{\hat{\mathbf{x}}_i
\cr\hat{\mathbf{y}}_i\cr\hat{\mathbf{z}}_i}\right].\label{matrix2}
\end{equation}
In Eq.~(\ref{matrix2}), we assume that terms equivalent to or higher than $(k_{x_i}/k)^2$, $(k_{y_i}/k)^2$ are null.

After we get the expression of electric field in $O_i-X_iY_iZ_i$ coordinate, the reflected and transmitted fields can be easily obtained by multiplying the Fresnel coefficients. We define the amplitude reflection coefficients of the main Fourier component as $r_p,~r_s$, the amplitude transmission coefficients of the main Fourier component as $t_p,~t_s$, where $p$ and $s$ denote $p$-polarized and $s$-polarized state. We also define the amplitude reflection coefficients of an arbitrary plane wave as $r_p^{'},~r_s^{'}$, the amplitude transmission coefficients of an arbitrary plane wave as $t_p^{'},~t_s^{'}$. To simplify the Fresnel
coefficients of an arbitrary wave component, we expand them around the central incident angle $\theta_i$ in series of $k_{x_i}/k$ and $k_{y_i}/k$, and retain the first order term. Thus, $r_{p,s}^{'}=r_{p,s}+\frac{\partial r_{p,s}}{\partial\theta_i}\frac{k_{x_i}}{k}$, $t_{p,s}^{'}=t_{p,s}+\frac{\partial t_{p,s}}{\partial\theta_i}\frac{k_{x_i}}{k}$. Note that, the reflected and transmitted electric fields are still presented in $O_r-X_rY_rZ_r$ and $O_t-X_tY_tZ_t$ coordinates. Hence, we need to transform them back to the $o_r-x_ry_rz_r$ and $o_t-x_ty_tz_t$ coordinates. The transformation matrixes are
\begin{eqnarray}
\left[\matrix{\hat{\mathbf{x}}_r\cr\hat{\mathbf{y}}_r\cr
\hat{\mathbf{z}}_r}\right]&=&\left[\matrix{1&\frac{k_{y_i}\cot\theta_i}{k}&-\frac{k_{x_i}}{k}\cr
-\frac{k_{y_i}\cot\theta_i}{k}&1&\frac{k_{y_i}}{k}\cr
\frac{k_{x_i}}{k}&-\frac{k_{y_i}}{k}&1}\right]\left[\matrix{\hat{\mathbf{X}}_i
\cr\hat{\mathbf{Y}}_i\cr\hat{\mathbf{Z}}_i}\right],\nonumber\\
\left[\matrix{\hat{\mathbf{x}}_t\cr\hat{\mathbf{y}}_t\cr
\hat{\mathbf{z}}_t}\right]&=&\left[\matrix{1&\frac{-\eta
k_{y_i}\cot\theta_i}{k}&\frac{k_{x_i}}{n\eta k}\cr \frac{\eta
k_{y_i}\cot\theta_i}{k}&1&\frac{k_{y_i}}{n k}\cr
-\frac{k_{x_i}}{n\eta k}&-\frac{k_{y_i}}{n
k}&1}\right]\left[\matrix{\hat{\mathbf{X}}_i
\cr\hat{\mathbf{Y}}_i\cr\hat{\mathbf{Z}}_i}\right].\label{matrix3}
\end{eqnarray}
The detailed procedures are similar to the Eq.~(\ref{matrix2}). As mentioned previously, we eliminate terms equal to or higher than $(k_{x_i}/k)^2$~and~ $(k_{y_i}/k)^2$.

The Fresnel coefficients only reveal the amplitudes of reflected and transmitted vortex beams. To get the full expressions of reflected and transmitted vortex beams, we still need the phase matching conditions. For an arbitrary plane wave component, the boundary conditions~\cite{Jackson1999} require that in the plane $xoy$, $\exp[i\mathbf{k}_i\cdot\mathbf{r}]=\exp[i\mathbf{k}_r\cdot\mathbf{r}]=\exp[i\mathbf{k}_t\cdot\mathbf{r}]$, where $\mathbf{k}_i$, $\mathbf{k}_r$, and $\mathbf{k}_t$ are the wave vectors of incident, reflected, and transmitted beams, respectively. Therefore, the phase matching conditions are presented
like this:
\begin{eqnarray}
k_{x_r}&=&-k_{x_i},~k_{y_r}=k_{y_i},~k_{z_r}=k_{z_i};\nonumber\\
k_{x_t}&=&k_{x_i}/\eta,~k_{y_t}=k_{y_i},~k_{z_t}=nk_{z_i}.\label{boundaryconditions}
\end{eqnarray}
Combining the Fresnel coefficients and Eqs.~(\ref{incidentbeam}) and (\ref{matrix2})-(\ref{boundaryconditions}), we get the electric fields of reflected and transmitted beams:
\begin{eqnarray}
\tilde{\mathbf{E}}_r&=&\left[\alpha\left(r_p-\frac{\partial
r_p}{\partial
\theta_i}\frac{k_{x_r}}{k}\right)+\beta\left(r_s+r_p\right)\cot\theta_i\frac{k_{y_r}}{k}\right]\tilde{u}_r\hat{\mathbf{x}}_r\nonumber\\
&+&\left[\beta\left(r_s-\frac{\partial r_s}{\partial
\theta_i}\frac{k_{x_r}}{k}\right)-\alpha\left(r_s+r_p\right)\cot\theta_i\frac{k_{y_r}}{k}\right]\tilde{u}_r\hat{\mathbf{y}}_r\nonumber\\
&-&\frac{1}{k}\left(\alpha r_p k_{x_r}+\beta r_s
k_{y_r}\right)\tilde{u}_r\hat{\mathbf{z}}_r,\label{reflectedbeam}
\end{eqnarray}
\begin{eqnarray}
\tilde{\mathbf{E}}_t&=&\left[\alpha\left(t_p+\eta\frac{\partial
t_p}{\partial
\theta_i}\frac{k_{x_t}}{k}\right)+\beta\left(t_p-\eta t_s\right)\cot\theta_i\frac{k_{y_t}}{k}\right]\tilde{u}_t\hat{\mathbf{x}}_t\nonumber\\
&+&\left[\beta\left(t_s+\eta\frac{\partial t_s}{\partial
\theta_i}\frac{k_{x_t}}{k}\right)+\alpha\left(\eta t_p-t_s\right)\cot\theta_i\frac{k_{y_t}}{k}\right]\tilde{u}_t\hat{\mathbf{y}}_t\nonumber\\
&-&\frac{1}{nk}\left(\alpha t_p k_{x_t}+\beta t_s
k_{y_t}\right)\tilde{u}_t\hat{\mathbf{z}}_t,\label{transmittedbeam}
\end{eqnarray}
where
$\tilde{u}_{r,t}=\tilde{u}_i\left(\gamma_{r,t}k_{x_{r,t}},k_{y_{r,t}}\right)$, $\gamma_r=-1,~\gamma_t=\eta$. Note that the reflected and transmitted vortex beams experience extra phase shifts~\cite{Goodman1996} associated with diffraction while propagating, which are $\exp\left[ikz_r\left(1-\frac{k_{x_r}^2+k_{y_r}^2}{2k^2}\right)\right]~\text{and}~\exp\left[inkz_t\left(1-\frac{k_{x_t}^2+k_{y_t}^2}{2n^2k^2}\right)\right]$,
respectively.
\begin{figure}
\includegraphics[width=8cm]{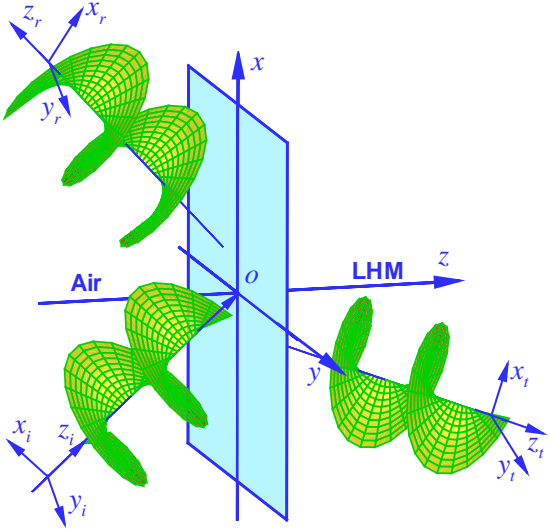}%
\caption{(Color online) The helical wavefronts of incident, reflected, and
transmitted beams. Incident, reflected, and transmitted vortex beams
propagate along $z_i,~z_r,~z_t$ axis, respectively. We choose $l=1$
for incident vortex beam. The incident, reflected, and transmitted
vortex beams have anti-clockwise, clockwise, and clockwise helical
wavefronts, respectively.}\label{wavefront}
\end{figure}

At this stage, we feel obliged to deliver several comments. First, the phase matching conditions, as demonstrated in Eq.~(\ref{boundaryconditions}), play a crucial role in this paper. For instance, $k_{x_r}=-k_{x_i}~\text{and}~k_{y_r}=k_{y_i}$ show that the wavefront of reflected vortex beam is reversed (shown in
Fig.~\ref{wavefront}), resulting in a reversed orbital angular momentum $-l\hbar$ per photon. $k_{z_t}=nk_{z_i}$ shows that the transmitted beam in LHM undergoes negative phase velocity, resulting in a reversed helical wavefront (shown in Fig.~\ref{wavefront}). The reversed wavefront in LHM does not lead to reversed orbital angular momentum, which is explained in Sec.~\ref{momenta}. The equation $k_{x_t}= k_{x_i}/\eta$ means the transmitted beam experiences a beam deformation in the cross section. Second, in the vicinity of Critical angle $\theta_C$ of TIR and Brewster angle $\theta_B$ ($|\theta_i-\theta_{B,C}|\leq \theta_0$, where $\theta_0=2/(kw_0)$ is the beam divergence angle), we should be cautious. For example, in the vicinity of Critical angle $|\theta_i-\theta_{C}|\leq \theta_0$), the first derivatives of $r_{p,s}$ are extremely large~\cite{Okuda2008}, meaning that Eqs.~(\ref{reflectedbeam})~and~(\ref{transmittedbeam}) should be revised to include terms higher than $(k_{x_i}/k)^2$ and $(k_{y_i}/k)^2$. If a p-polarized vortex beam impinges on the air-LHM interface with incident angle being in the vicinity of Brewster angle ($|\theta_i-\theta_{B}|\leq \theta_0$), the power of
reflected beam would be in the scale of $\theta_0^2$ (approximately $10^{-4}\sim 10^{-6}$ for paraxial beams) comparing with the incident beam. This constitutes the reason why the experiments~\cite{Dasgupta2006,Qin2009} fail to collect reliable data concerning the shift of p-polarized beams in the vicinity of Brewster angle. Third, although the beam shape and shift of reflected vortex beam are already clear~\cite{Fedoseyev2001,Bliokh2009,Merano2010}. The profile of transmitted vortex beam is far from clear-cut. We formulate a rigorous transmitted field in the appendix of our paper. By referencing this appendix, the beam shape of the transmitted vortex beam is easily acquired.

\section{Spatial and Angular Shifts}\label{shifts}
The paraxial wave equation is identical with two-dimensional Schr\"{o}dinger equation with $z$ replaced by $t$. Hence, the operator formalism proposed by Stoler~\cite{Stoler1981} has become a powerful tool in physical optics~\cite{Enk1994,Aiello2009,Bliokh2010}. In this section, we will apply the operator formalism to calculate the centroid of the reflected and transmitted beams. In momentum space, the transverse position operator is $i\partial_{\mathbf{k}_\bot}$. Therefore, the centroid of beams in a given plane $z=const$ is readily given by
\begin{equation}
\langle\mathbf{r}_{\perp}\rangle=\frac{\langle
G\tilde\mathbf{E}|i\partial_{\mathbf{k_\perp}}|G\tilde\mathbf{E}\rangle}{\langle
G\tilde\mathbf{E}|G\tilde\mathbf{E}\rangle},\label{beamcentroid_1}
\end{equation}
where $\mathbf{r}_{\perp}=x\hat{\mathbf{x}}+y\hat{\mathbf{y}}$, $\partial_{\mathbf{k_\perp}}=\frac{\partial}{\partial k_x}\hat{\mathbf{x}}+\frac{\partial}{\partial k_y}\hat{\mathbf{y}}$, the propagation operator $G=\exp\left[inkz-\frac{iz}{2nk}\left(k_x^2+k_y^2\right)\right]$.
The above equation can be easily formulated into the following one:
\begin{equation}
\langle\mathbf{r}_{\perp}\rangle=\frac{\langle
\tilde\mathbf{E}|i\partial_{\mathbf{k_\perp}}|\tilde\mathbf{E}\rangle}{\langle
\tilde\mathbf{E}|\tilde\mathbf{E}\rangle}+\frac{z}{nk}\frac{\langle
\tilde\mathbf{E}|\mathbf{k}_\perp|\tilde\mathbf{E}\rangle}{\langle
\tilde\mathbf{E}|\tilde\mathbf{E}\rangle},\label{beamcentroid_2}
\end{equation}
where $k$ is the wave number in vacuum and $n$ is the refractive index. The first and second terms of Eq.~(\ref{beamcentroid_2}) are the spatial and angular shifts, which are independent and dependent on $z$, respectively. From the second term of Eq.~(\ref{beamcentroid_2}), we can easily find that the negative
refractive index $n$ results in reversed angular shifts in LHM~\cite{Luo2009}. For the reflected and transmitted beams, the GH and IF shifts can be expressed as: $\langle x_\tau\rangle=\Delta x_\tau+z_\tau\Delta \theta_{x_\tau}$ and $\langle y_\tau\rangle=\Delta y_\tau+z_\tau\Delta\theta_{y_\tau}$, where $\tau=r,t$.

We first calculate the spatial and angular shifts of reflected beams. Substituting Eq.~(\ref{reflectedbeam}) into Eq.~(\ref{beamcentroid_2}), we get the spatial and angular shifts of reflected beams
\begin{eqnarray}
\Delta x_r&=&\frac{\chi
l\cot\theta_i\left(\left|r_s\right|^2-\left|r_p\right|^2\right)}{2k\left(\left|\alpha
r_p\right|^2+\left|\beta r_s\right|^2\right)}\nonumber\\
&&+\frac{\left|\alpha r_p\right|^2\frac{\partial
\phi_{r_p}}{\partial \theta_i}+\left|\beta
r_s\right|^2\frac{\partial \phi_{r_s}}{\partial
\theta_i}}{k\left(\left|\alpha r_p\right|^2+\left|\beta
r_s\right|^2\right)},\label{x_r}\\
\Delta y_r&=&\frac{-l\left(\left|\alpha
\right|^2\left|r_p\right|\frac{\partial\left|r_p\right|}{\partial
\theta_i}+\left|\beta\right|^2\left|r_s\right|\frac{\partial
\left|r_s\right|}{\partial \theta_i}\right)}{k\left(\left|\alpha
r_p\right|^2+\left|\beta r_s\right|^2\right)}\nonumber\\
&&-\frac{\cot \theta_i}{2k\left(\left|\alpha
r_p\right|^2+\left|\beta
r_s\right|^2\right)}\bigg\{2\chi|r_s||r_p|\sin(\phi_{r_s}-\phi_{r_p})\nonumber\\
&&+\sigma\big[|r_p|^2+|r_s|^2+2|r_p||r_s|\cos(\phi_{r_s}-\phi_{r_p})\big]\bigg\},\label{y_r}\\
\Delta \theta_{x_r}&=&\frac{-(|l|+1)\left(\left|\alpha
\right|^2\left|r_p\right|\frac{\partial\left|r_p\right|}{\partial
\theta_i}+\left|\beta\right|^2\left|r_s\right|\frac{\partial
\left|r_s\right|}{\partial \theta_i}\right)}{kz_R\left(\left|\alpha
r_p\right|^2+\left|\beta r_s\right|^2\right)},\label{x_theta_r}\\
\Delta
\theta_{y_r}&=&\frac{\chi(|l|+1)\cot\theta_i\left(\left|r_p\right|^2-\left|r_s\right|^2\right)}{2kz_R\left(\left|\alpha
r_p\right|^2+\left|\beta r_s\right|^2\right)},\label{y_theta_r}
\end{eqnarray}
where $r_{p,s}=|r_{p,s}|\exp[{i\phi_{r_{p,s}}}]$, $\sigma=2\text{Im}[\alpha^*\beta]$, $\chi=2\text{Re}[\alpha^*\beta]$, $z_R=kw_0^2/2$ is the Rayleigh
length. When evaluating the energy term $\langle \tilde\mathbf{E}|\tilde\mathbf{E}\rangle$, we discard the cross-polarization terms (terms proportional to $k_{x_r}/k$ or $k_{y_r}/k$) and $z_r$-component electric field $\tilde{E}_{z_r}$, since their average energy density is proportional to $\theta_0^2$.
The first terms of Eqs.~(\ref{x_r}) and (\ref{y_r}) are vortex-induced spatial GH and IF shifts. They were initially proposed by Bliokh~\cite{Bliokh2009} and
Fedoseyev~\cite{Fedoseyev2001}, respectively. Experimental demonstrations were accomplished by Merano~\cite{Merano2010} and Dasgupta~\cite{Dasgupta2006} at air-glass interface. Note that the vortex-induced spatial GH shift only exists in mixed linearly polarized beams (polarization vector is oblique to the incident
plane), while the vortex-induced spatial IF shift occurs in arbitrary polarized state. The second term of Eq.~(\ref{x_r}) coincides with the well-known Artmann formula~\cite{Artmann1948}. The second term of Eq.~(\ref{y_r}) is the spin-dependent IF shift~\cite{Schilling1965,Bliokh2007,Aiello2008}. In partial
reflection and TIR, it turns into the Bliokh formula~\cite{Bliokh2007} and Schillings formula~\cite{Schilling1965}, respectively. Equations~(\ref{x_theta_r})~and~(\ref{y_theta_r}) are the angular shifts. In general, they are proportional to $\theta_0^2$. For p-polarized beams, these angular shifts could be remarkably magnified through Brewster resonance~\cite{Merano2009} (finally in the scale of $\theta_0$ near the Brewster angle). In these cases,
the cross-polarization terms are not negligible when we calculate the energy intensity.
\begin{figure}
\includegraphics[width=8cm]{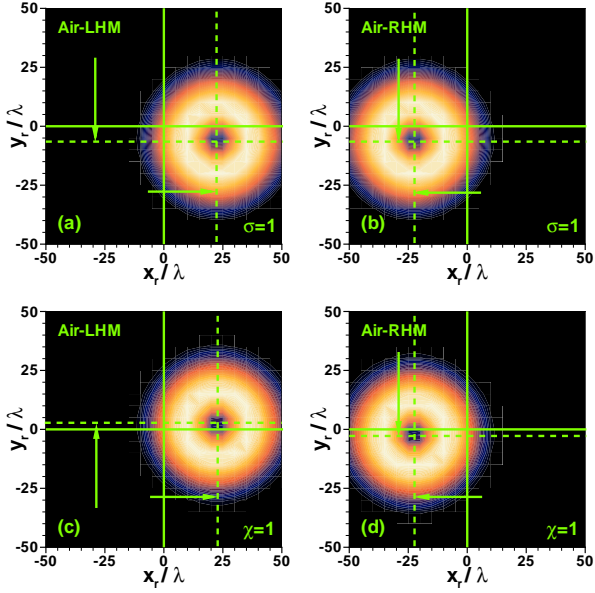}%
\caption{(Color online) The GH and IF shifts in TIR. The parameters are: incident
angle $\theta_i=\pi/4$, refractive index of LHM $n_L=-1.515$,
refractive index of RHM $n_R=1.515$, vortex charge $l=1$,
propagation distance $z=z_R$, beam waist $w_0=20\lambda$, where
$\lambda$ is the wavelength in vacuum. To make the shifts more
noticeable, we amplify them by 20 times. In the first row, the
incident beam is left-circularly polarized.  The actual beam shifts
of Fig.~\ref{totalreflection}(a) are $\langle
x_r^L\rangle=1.129\lambda$ and $\langle y_r^L\rangle=-0.282\lambda$.
The actual beam shifts of Fig.~\ref{totalreflection}(b) are $\langle
x_r^R\rangle=-1.129\lambda$ and $\langle
y_r^R\rangle=-0.282\lambda$. In the second row, the incident beam is
mixed linearly polarized with $\chi=1$. The actual beam shifts of
Fig.~\ref{totalreflection}(c) are $\langle
x_r^L\rangle=1.129\lambda$ and $\langle y_r^L\rangle=0.101\lambda$.
The actual beam shifts of Fig.~\ref{totalreflection}(d) are $\langle
x_r^R\rangle=-1.129\lambda$ and $\langle
y_r^R\rangle=-0.101\lambda$. }\label{totalreflection}
\end{figure}

A careful assessment of Eqs.~(\ref{x_r})-(\ref{y_theta_r}) indicates that, in partial reflection region, the spatial and angular GH and IF shifts at loss-free air-LHM and air-RHM interfaces are identical. In TIR, shifts are different, however. In this case, the vortex-induced shifts and angular shifts are null. The phase $\phi_{r_{p,s}}$ has the same magnitude but the opposite sign for RHM and LHM, resulting in a negative GH shift~\cite{Berman2002,Shadrivov2003} in LHM. The IF shifts for circularly polarized beams and mixed linearly polarized beams are $-\sigma\cot\theta_i[1+\cos(\phi_{r_s}-\phi_{r_p})]/k$ and $-\chi\cot\theta_i\sin(\phi_{r_s}-\phi_{r_p})/k$, respectively. Here, we theoretically predict that owing to the phase reversion, the IF shift in total reflection region would also be reversed at air-LHM interface when the incident beam is in mixed linearly polarized state. This prediction is illustrated in
Fig.~\ref{totalreflection}. With the refractive index $n_R=-n_L=1.515$, we choose incident $\theta_i=\pi/4$ to avoid the deformation of reflected beams~\cite{Okuda2008}. Figure~\ref{totalreflection}(a) and Fig.~\ref{totalreflection}(b) show that the IF shift in TIR remains unreversed when the incident beam is in circularly polarized state. Figure~\ref{totalreflection}(c) and Fig.~\ref{totalreflection}(d) suggest that the IF shift in TIR is reversed when the incident beam is in mixed linearly polarized state.
\begin{figure}
\includegraphics[width=8cm]{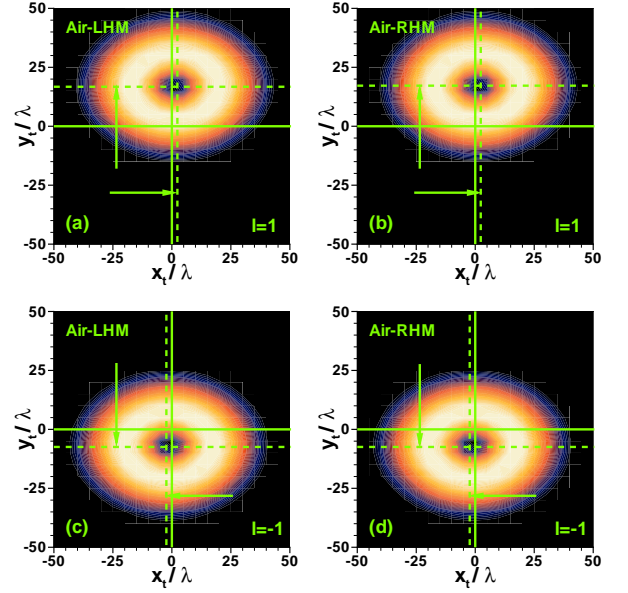}%
\caption{(Color online) The spatial GH and IF shifts of transmitted vortex beams at
air-LHM and air-RHM interfaces. The parameters are: incident angle
$\theta_i=\pi/4$, refractive index of LHM $n_L=-1.515$, refractive
index of RHM $n_R=1.515$, polarization state $\sigma=1/2$,
$\chi=1/2$, propagation distance $z=z_R$, beam waist
$w_0=20\lambda$. To make the shifts more noticeable, we amplify them
by 200 times. In the first row, the vortex charge is $l=1$. The
actual beam shifts of Fig.~\ref{partialtransmission}(a) are $\Delta
x_t^L=0.006\lambda$ and $\Delta y_t^L=0.085\lambda$.
Figure~\ref{partialtransmission}(b) has the same shifts with
Fig.~\ref{partialtransmission}(a). In the second row, the vortex
charge is $l=-1$. The actual beam shifts of
Fig.~\ref{partialtransmission}(c) are $\Delta x_t^L=-0.006\lambda$
and $\Delta y_t^L=-0.033\lambda$.
Figure~\ref{partialtransmission}(d) has the same shifts with
Fig.~\ref{partialtransmission}(c). }\label{partialtransmission}
\end{figure}
\begin{figure}
\includegraphics[width=8cm]{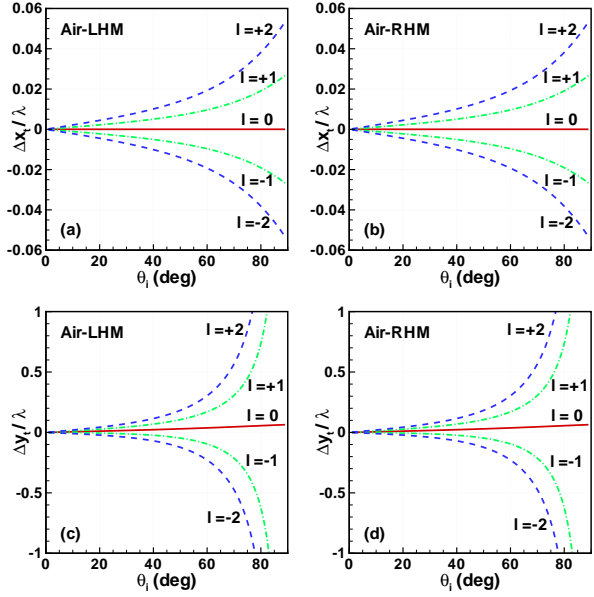}%
\caption{(Color online) The spatial GH and IF shifts of transmitted vortex beams at
air-LHM and air-RHM interfaces. The parameters are: the refractive
index of LHM $n_L=-1.515$, the refractive index of RHM $n_R=1.515$,
polarization state $\sigma=1/2$, $\chi=1/2$. The spatial shifts are
presented in the scale of $\lambda$. }\label{spatialshifts}
\end{figure}

Now, we begin to study the shifts of transmitted vortex in LHM. We substitute Eq.~(\ref{transmittedbeam}) into Eq.~(\ref{beamcentroid_2}) and get the spatial and angular shifts of transmitted beams
\begin{eqnarray} \Delta x_t&=&\frac{\chi\eta
l\cot\theta_i\left(\left|t_p\right|^2-\left|t_s\right|^2\right)}{2k\left(\left|\alpha
t_p\right|^2+\left|\beta t_s\right|^2\right)}\nonumber\\
&&-\frac{\eta\left(\left|\alpha t_p\right|^2\frac{\partial
\phi_{t_p}}{\partial \theta_i}+\left|\beta
t_s\right|^2\frac{\partial \phi_{t_s}}{\partial
\theta_i}\right)}{k\left(\left|\alpha t_p\right|^2+\left|\beta
t_s\right|^2\right)},\label{x_t}\\
\Delta y_t&=&\frac{-l\left(\left|\alpha
\right|^2\left|t_p\right|\frac{\partial\left|t_p\right|}{\partial
\theta_i}+\left|\beta\right|^2\left|t_s\right|\frac{\partial
\left|t_s\right|}{\partial \theta_i}\right)}{k\left(\left|\alpha
t_p\right|^2+\left|\beta t_s\right|^2\right)}\nonumber\\
&&+\frac{\cot \theta_i}{2k\left(\left|\alpha
t_p\right|^2+\left|\beta
t_s\right|^2\right)}\bigg\{2\chi\eta|t_s||t_p|\sin(\phi_{t_s}-\phi_{t_p})\nonumber\\
&&-\sigma\big[|t_p|^2+|t_s|^2-2\eta|t_p||t_s|\cos(\phi_{t_s}-\phi_{t_p})\big]\bigg\},\label{y_t}\\
 \Delta \theta_{x_t}&=&\frac{\eta(|l|+1)\left(\left|\alpha
\right|^2\left|t_p\right|\frac{\partial\left|t_p\right|}{\partial
\theta_i}+\left|\beta\right|^2\left|t_s\right|\frac{\partial
\left|t_s\right|}{\partial
\theta_i}\right)}{kz_{R_x}\left(\left|\alpha
t_p\right|^2+\left|\beta t_s\right|^2\right)},\label{x_theta_t}\\
\Delta
\theta_{y_t}&=&\frac{\chi(|l|+1)\cot\theta_i\left(\left|t_p\right|^2-\left|t_s\right|^2\right)}{2kz_{R_y}\left(\left|\alpha
t_p\right|^2+\left|\beta t_s\right|^2\right)},\label{y_theta_t}
\end{eqnarray}
where $t_{p,s}=|t_{p,s}|\exp[{i\phi_{t_{p,s}}}]$, $z_{R_x}=nk\eta^2w_0^2/2$ and $z_{R_y}=nkw_0^2/2$ is the Rayleigh length along $x_t$ and $y_t$ axis. These Rayleigh lengths of transmitted vortex beam in LHM are both negative owing to negative phase velocity~\cite{Luo2006}. The first terms of Eqs.~(\ref{x_t})
and (\ref{y_t}) are vortex-induced spatial GH and IF shifts. Until now, no experiments have been reported on these vortex-induced GH and IF shifts. The vortex-induced GH shift of transmitted beam occurs when the incident beam is in mixed linearly polarized state. The vortex-induced IF shift, however, exists in any polarization state. The second term of Eq.~(\ref{x_t}) is the phase-dependent GH shift. In lossy media, this part is not negligible. The second term
of Eq.~(\ref{y_t}) is the spin-dependent IF shift. If the LHM is loss-free, this term degenerates into Bliokh formula~\cite{Bliokh2007}. A special case is ``total transmission", where $n=-1$. In this case, $t_p=t_s=1$, we get $\langle x_t\rangle=\langle y_t\rangle=0$.
\begin{figure}
\includegraphics[width=8cm]{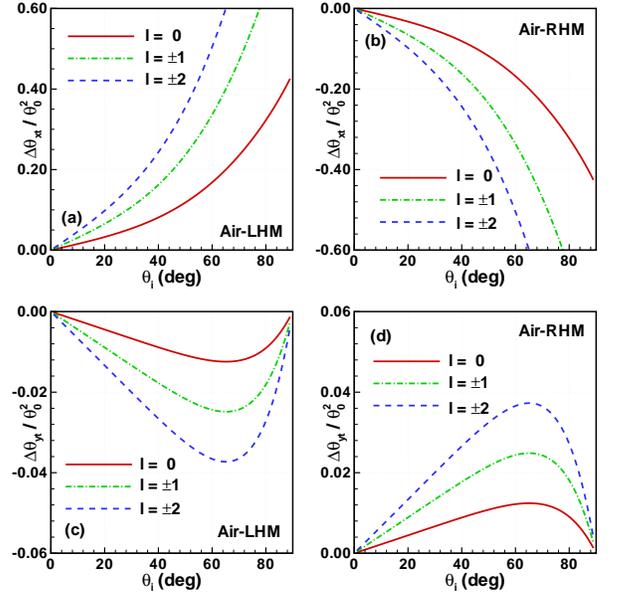}%
\caption{(Color online) The angular GH and IF shifts of transmitted vortex beams at
air-LHM and air-RHM interfaces. The parameters are: the refractive
index of LHM $n_L=-1.515$, the refractive index of RHM $n_R=1.515$,
polarization state $\sigma=1/2$, $\chi=1/2$. The angular shifts are
presented in the scale of $\theta_0^2$. }\label{angularshifts}
\end{figure}

We closely examine Eqs.~(\ref{x_t})-(\ref{y_theta_t}) and find that spatial GH and IF shifts of partial transmitted beams remain unreversed at air-LHM interface compared with air-RHM interface. The angular shifts, however, are reversed. Figure~\ref{partialtransmission} shows the spatial GH and IF shifts of transmitted beams at air-LHM and air-RHM interfaces. The beam profiles are drawn according to Eqs.~(\ref{transmittedbeam_position_space_1}), (\ref{vortex_1}) and
(\ref{vortex_-1}). We can easily find that the transmitted beam is stretched along $x_t$ axis, which is a pure geometrical phenomenon. We assume that the LHM is low-loss or loss-free. This assumption is reasonable considering state-of-the-art micro/nano manufacturing technology. For example, by incorporating gain media into fishnet structure~\cite{Xiao2010} or exploiting second-order magnetic resonance of the fishnet structure~\cite{Garcia2011}, we can obtain low-loss, 3-D, polarization-independent LHM in the visible spectral range. Based on this assumption, the phase-dependent term of spatial GH shift vanishes and the spatial GH shift only depends on vortex. This argument is confirmed by Fig.~\ref{partialtransmission}(a) and Fig.~\ref{partialtransmission}(c), where the vortex charge is $l=1$ and $l=-1$, respectively. They indicate that when the vortex charge is reversed, the spatial GH shift is also reversed. The spatial IF shift is both vortex and spin-dependent. Therefore, when the vortex charge is flipped, the absolute value of spatial IF shift is altered [but not flipped, see Fig.~\ref{partialtransmission}(a) and Fig.~\ref{partialtransmission}(c)]. In Fig.~\ref{spatialshifts}, we demonstrates how incident angle impact the spatial GH and IF shifts. When the incident angle increases, both GH and IF shifts increase. The IF shift fails to converge when $\theta_i$ approaches to
$90^{\circ}$. We note that in this case, the energy transmission coefficient is zero. Hence, the spatial shifts of transmitted beams are meaningless. Figure~\ref{spatialshifts} also suggests that both spatial GH and IF can be enhanced by raising the vortex charge.

We proceed to analyze the angular GH and IF shifts. Figure~\ref{angularshifts} demonstrates how the incident angle affects the angular shifts. Figure ~\ref{angularshifts}(a) and Fig.~\ref{angularshifts}(b) show the angular GH shift at air-LHM and air-RHM interfaces, respectively. They indicate that the angular GH shift is reversed in air-LHM. Figure ~\ref{angularshifts}(c) and Fig.~\ref{angularshifts}(d) show the angular IF shift at air-LHM and air-RHM interfaces, respectively. They indicate that the angular IF shift is reversed in air-LHM. For angular IF shift in air-LHM, there exists a minimum point as incident increases from $0^{\circ}$ to $90^{\circ}$. These angular shifts increase proportionally as we raise the absolute value of vortex charge.

\section{Linear and Angular Momenta}\label{momenta}
In this section, we start to analyze the linear and angular momenta of reflected and transmitted vortex beams. In general, the linear momentum density can be divided into orbital part and spin part~\cite{Berry2009}, which reads
\begin{equation}
\mathbf{p}=\text{Im}[\mathbf{E}^*\times(\nabla)\mathbf{E}]+\frac{1}{2}\text{Im}[\nabla\times(\mathbf{E}^*\times\mathbf{E})],\label{momentumdensity}
\end{equation}
where the first term is orbital momentum density $\mathbf{p}^{o}$, the second term is spin momentum density $\mathbf{p}^{s}$, $\mathbf{E}$ is the electric field in position space. In momentum space, we can write Eq.~(\ref{momentumdensity}) into
\begin{equation}\label{momentum_density}
\mathbf{p}=|G\tilde{\mathbf{E}}|^2\mathbf{k}+(k_x\hat{\mathbf{y}}-k_y\hat{\mathbf{x}})(G^*\tilde{E}_x^*G\tilde{E}_y-G\tilde{E}_x
G^*\tilde{E}_y^*).
\end{equation}
Therefore, the momentum per unit length is
\begin{equation}\label{linear_momentum}
\mathbf{P}=\langle\tilde\mathbf{E}|\mathbf{k}|\tilde\mathbf{E}\rangle+\langle\tilde{E}_x|k_x|\tilde{E}_y\rangle\hat{\mathbf{y}}-\langle\tilde{E_y}|k_y|\tilde{E_x}\rangle\hat{\mathbf{x}},
\end{equation}
where the first term is orbital momentum $\mathbf{P}^{O}$, the second and third terms are spin momentum $\mathbf{P}^{S}$. From the above equation, we find that $P_z\propto nk$, which means $P_z$ in LHM is opposite to $P_z$ in RHM. For both paraxial and nonparaxial beams, the spin part makes no contribution to the linear momentum~\cite{Li2009,Bliokh2010}: $\mathbf{P}^{S}=0$.

In position space, the angular momentum density denotes $\mathbf{j}=\mathbf{r}\times\mathbf{p}$. In momentum space, we replace $\mathbf{r}$ with operator
$i\partial_{\mathbf{k}_\bot}+z\hat{\mathbf{z}}$ and get the angular momentum
\begin{eqnarray}\label{angular_momentum}
\mathbf{J}&=&\langle
G\tilde{\mathbf{E}}|(i\partial_{\mathbf{k}_\bot}+z\hat{\mathbf{z}})\times\mathbf{k}|G\tilde{\mathbf{E}}\rangle\nonumber\\
&&+\big[\langle G\tilde{E_y}|i(\partial_{k_x}k_x+\partial_{k_y}k_y)|G\tilde{E_x}\rangle\nonumber\\
&&-\langle
G\tilde{E_x}|i(\partial_{k_x}k_x+\partial_{k_y}k_y)|G\tilde{E_y}\rangle\big]\hat{\mathbf{z}},
\end{eqnarray}
where the first term is orbital angular momentum $\mathbf{J}^O$, the second and third terms are spin angular momentum $\mathbf{J}^S$. Note that the transverse spin angular momenta are null, which means $J_x^S=J_y^S=0$. We substitute the propagation operator $G=\exp\left[inkz-\frac{iz}{2nk}\left(k_x^2+k_y^2\right)\right]$
into Eq.~(\ref{angular_momentum}) and get the orbital and spin angular momentum~\cite{Li2009,Aiello2009,Bliokh2010}
\begin{eqnarray}
\mathbf{J}^O&=&\langle\tilde{\mathbf{E}}|ink\partial_{k_y}|\tilde{\mathbf{E}}\rangle\hat{\mathbf{x}}+\langle\tilde{\mathbf{E}}|-ink\partial_{k_x}|\tilde{\mathbf{E}}\rangle\hat{\mathbf{y}}\nonumber\\
&&+\langle\tilde{\mathbf{E}}|i(k_y\partial_{k_x}-k_x\partial_{k_y})|\tilde{\mathbf{E}}\rangle\hat{\mathbf{z}},\label{orbital_angular momentum}\\
\mathbf{J}^S&=&\big[\langle\tilde{E_y}|i(\partial_{k_x}k_x+\partial_{k_y}k_y)|\tilde{E_x}\rangle\nonumber\\
&&-\langle\tilde{E_x}|i(\partial_{k_x}k_x+\partial_{k_y}k_y)|\tilde{E_y}\rangle\big]\hat{\mathbf{z}}.\label{spin_angular_momentum}
\end{eqnarray}
We find that the transverse orbital angular momenta $J_x^O$ and $J_y^O$ are proportional to $nk$, which means they have opposite directions in LHM compared with the counterparts in RHM. Equations~(\ref{orbital_angular momentum}) and (\ref{spin_angular_momentum}) also indicate that the $z$-component orbital angular momentum $J_z^O$ and spin angular momentum $J_z^S$ have the same directions in LHM and RHM. Comparing Eq.~(\ref{beamcentroid_2}) with Eqs.~(\ref{linear_momentum}) and (\ref{orbital_angular momentum}), we can easily get the following relations
\begin{equation}
\Delta x=-\frac{J_y^O}{P_z^O}, \Delta y=\frac{J_x^O}{P_z^O};
\Delta\theta_x=\frac{P_x^O}{P_z^O},\Delta\theta
_y=\frac{P_y^O}{P_z^O},\label{sptail_and_angular_shfits}
\end{equation}
where $\Delta x$ and $\Delta y$ are spatial shifts, $\Delta\theta x$
and $\Delta\theta y$ are angular shifts.

We first offer a qualitative explanation on why spatial shifts are unreversed in LHM but angular shifts are reversed in LHM. Equation~(\ref{sptail_and_angular_shfits}) indicates that spatial shifts have no direct relations with the $z$-component orbital angular momentum $J_z^O$ and spin angular momentum $J_z^S$. They are dependent on the transverse angular momenta $\mathbf{J}_\bot^O$ and $z$-component linear momentum $P_z^O$~\cite{Aiello2009}. Since the directions of $\mathbf{J}_\bot^O$ and $P_z^O$ are both reversed in LHM, the spatial shifts $\Delta \mathbf{r}_\bot$ would remain unreversed as a result. The angular shifts, however, solely depend on linear momentum. The reason why angular shifts are reversed in LHM is that the transverse linear momentum $\mathbf{P}_\bot^O$ is unreversed but the $z$-component linear momentum $P_z^O$ is reversed.

From now on, we will perform a quantitative analysis on the linear and angular momenta and confirm the momentum conservation laws. Therefore, we assume that the LHM is loss-free and the amplitude reflection coefficients $r_{p,s}$ and amplitude transmission coefficients $t_{p,s}$ are real variables. By substituting
Eqs.~(\ref{incidentbeam}), (\ref{reflectedbeam}), and (\ref{transmittedbeam}) into Eq.~(\ref{linear_momentum}), we obtain the linear momenta of incident, reflected, and transmitted beams
\begin{eqnarray}
\mathbf{P}_i&=&k\hat{\mathbf{z}}_i,\nonumber\\
\mathbf{P}_r&=&\frac{-(|l|+1)}{2z_R}\frac{\partial Q_r}{\partial
\theta_i}\hat{\mathbf{x}}_r+\frac{\chi(|l|+1)}{2z_R}(r_p^2-r_s^2)\cot\theta_i\hat{\mathbf{y}}_r\nonumber\\
&&+kQ_r\hat{\mathbf{z}}_r,\nonumber\\
\mathbf{P}_t&=&\frac{(|l|+1)}{2\eta^2z_R}\frac{\partial T}{\partial\theta_i}\hat{\mathbf{x}}_t+\frac{\chi(|l|+1)}{2\eta z_R}(t_p^2-t_s^2)\cot\theta_i\hat{\mathbf{y}}_t\nonumber\\
&&+\frac{nkT}{\eta}\hat{\mathbf{z}}_t,
\end{eqnarray}
where $Q_r=|\alpha r_p|^2+|\beta r_s|^2$ is the energy reflection coefficient, $T=|\alpha t_p|^2+|\beta t_s|^2$. The energy transmission coefficient is $Q_t=|n|\eta T$. Note that the linear momenta of incident, reflected, and transmitted beams are presented in three different coordinate systems
$o_{i,r,t}-x_{i,r,t}y_{i,r,t}z_{i,r,t}$. We can verify that the linear momenta along $x$ and $y$ axes satisfy the conservation law~\cite{Fedoseyev2009,Fedoseyev2011}
\begin{eqnarray}
P_{z_i}\sin\theta_i&=&P_{z_r}\sin\theta_i-P_{x_r}\cos\theta_i\nonumber\\
&&+|n|\eta^2(P_{x_t}\cos\theta_t-P_{z_t}\sin\theta_t),\nonumber\\
P_{y_i}&=&P_{y_r}+|n|\eta^2P_{y_t}=0. \label{linear_momenta_conservation}
\end{eqnarray}
For incident beam, there is no transverse linear momentum. To satisfy the linear momentum conservation law, transverse linear momenta $P_{x_r}$, $P_{y_r}$, $P_{x_t}$, $P_{y_t}$ are produced. These transverse linear momenta are responsible for angular shifts.

We proceed to analyze the angular momenta. By substituting Eqs.~(\ref{incidentbeam}), (\ref{reflectedbeam}), and (\ref{transmittedbeam}) into Eqs.~(\ref{orbital_angular momentum}) and (\ref{spin_angular_momentum}), we get the angular momenta of incident, reflected, and transmitted beams.
\begin{eqnarray}
\mathbf{J}_i&=&(l+\sigma)\hat{\mathbf{z}_i},\nonumber\\
\mathbf{J}_r&=&\bigg[-\frac{l}{2}\frac{\partial
Q_r}{\partial \theta_i}-\frac{\sigma}{2}(r_p+r_s)^2\cot\theta_i\bigg]\hat{\mathbf{x}}_r\nonumber\\
&&+\frac{\chi l}{2}(r_p^2-r_s^2)\cot\theta_i\hat{\mathbf{y}}_r+(-l
Q_r+\sigma r_pr_s)\hat{\mathbf{z}}_r,\nonumber\\
\mathbf{J}_t&=&\bigg[-\frac{nl}{2\eta}\frac{\partial
T}{\partial\theta_i}-\frac{n\sigma}{2\eta}(t_p^2+t_s^2-2\eta
t_pt_s)\cot\theta_i\bigg]\hat{\mathbf{x}}_t\nonumber\\
&&+\frac{n\chi
l}{2}(t_s^2-t_p^2)\cot\theta_i\hat{\mathbf{y}}_t+\bigg[\frac{l(1+\eta^2)T}{2\eta^2}+\frac{\sigma
t_pt_s}{\eta}\bigg]\hat{\mathbf{z}}_t,\nonumber\\\label{angular_momenta}
\end{eqnarray}
where the spin angular momenta of incident, reflected, and transmitted beams are $\sigma$, $\sigma r_pr_s$, and $\sigma t_pt_s\eta^{-1}$, respectively. For each individual photon, the angular momenta of incident, reflected, and transmitted beams are $(l+\sigma)\hbar$, $(-l+\sigma r_pr_s/Q_r)\hbar$,
$[l(\eta+\eta^{-1})/2+\sigma t_pt_s/T]\hbar$, respectively. Though the orbital and spin momenta of reflected photon and transmitted photon depend on the absolute value refractive index, they are independent on the sign of refractive index. The $z$-component angular momenta satisfy the conservation law:
\begin{eqnarray}
J_{z_i}\cos\theta_i&=&-J_{z_r}\cos\theta_i-J_{x_r}\sin\theta_i\nonumber\\
&&+|n|\eta^2(J_{x_t}\sin\theta_t+J_{z_t}\cos\theta_t).\label{angular_momenta_conservation}
\end{eqnarray}
From Eqs.~(\ref{angular_momenta}) and
(\ref{angular_momenta_conservation}), we can infer that there are two types of momentum conversions. The first type is spin-orbit conversion. The $z$-component spin angular momentum of incident beam $\sigma\cos\theta_i$ converts into transverse angular momenta $J_{x_r}^O$ and $J_{x_t}^O$, resulting in spin-dependent IF shifts. The second type is orbit-orbit conversion. The $z$-component orbital angular momentum of incident beam $l\cos\theta_i$ converts into transverse angular momenta $J_{x_r}^O$ and $J_{x_t}^O$, resulting in vortex-induced IF shifts. The vortex-induced GH shifts, though relate to $J_{y_{r,t}}^O$, are not governed by angular momenta conservation law.

In TIR, the linear and angular momenta of incident and reflected beams are
\begin{eqnarray}
\mathbf{P}_i&=&k\hat{\mathbf{z}}_i,~~\mathbf{P}_r=k\hat{\mathbf{z}}_r;\nonumber\\
\mathbf{J}_i&=&(l+\sigma)\hat{\mathbf{z}_i},\nonumber\\
\mathbf{J}_r&=&-\cot\theta_i\big[\sigma+\sigma\cos(\phi_{r_s}-\phi_{r_p})+\chi\sin(\phi_{r_s}-\phi_{r_p})\big]\hat{\mathbf{x}}_r\nonumber\\
&&-\left(|\alpha|^2\frac{\partial \phi_{r_p}}{\partial
\theta_i}+|\beta|^2\frac{\partial \phi_{r_s}}{\partial
\theta_i}\right)\hat{\mathbf{y}}_r\nonumber\\
&&+\big[-l+\sigma\cos(\phi_{r_s}-\phi_{r_p})+\chi\sin(\phi_{r_s}-\phi_{r_p})\big]\hat{\mathbf{z}}_r,\nonumber\\
\end{eqnarray}
We can easily verify that they fulfil the conservation law
\begin{eqnarray}
P_{z_i}\sin\theta_i&=&P_{z_r}\sin\theta_i,\nonumber\\
J_{z_i}\cos\theta_i&=&-J_{z_r}\cos\theta_i-J_{x_r}\sin\theta_i.
\end{eqnarray}

At this point, we would like to add three comments. First, it's worth noting that we adopted the Minkowski momentum in Eqs.~(\ref{linear_momenta_conservation}) and (\ref{angular_momenta_conservation}) in this paper. Although the Abraham-Minkowski dilemma has been solved~\cite{Barnett2010,Milonni2010,Kemp2011}, why Minkowski momentum is a proper form in this study, we believe, is an interesting problem worth further investigation. Second, two important papers~\cite{Bekshaev2012,Aiello2012} were published while our paper was being peer-reviewed. One paper~\cite{Bekshaev2012} adopted the real-space approach and revealed the role of longitudinal field $E_z$ in IF shifts for the first time. We think additional analysis on the connection between longitudinal field and spin-orbit conversion can also be carried out in the momentum space. The other paper~\cite{Aiello2012} unambiguously separated the effects of beam shape and other parameters (such as polarization, the property of the interface) on GH and IF shifts, which has long escaped researchers' attention over the past years. But the discussions are confined to reflected vortex beam. Generalization to transmitted vortex beam, we think, remains quite challenging. Third, Owing to the close similarity between light beam and matter waves, scientists have found that electron beams can also possess orbital angular momentum by passing through a spiral phase plate~\cite{Uchida2010} or nanofabricated diffraction hologram~\cite{Verbeeck2010}. In this regard, we hopefully predict that vortex electron beam might also experience vortex-induced shifts in a potential well. By properly designing the potential well and taking advantage of quantum weak measurements or other measuring technology, we may even observe the vortex-induced beam shifts of electron beam in experiment.

\section{Conclusions}
In conclusion, we have derived the reflected
and transmitted fields of vortex beam at air-LHM interface via angular spectrum method. By using this method, we have managed to get the formulas of spatial GH
shifts, spatial IF shifts, angular GH shifts, and angular IF shifts. These formulas suggest that the spatial GH and IF shifts remain unreversed at air-LHM interface compared with air-RHM interface. By raising the vortex charge, the spatial shifts can be remarkably enhanced. In TIR, apart from reversed GH shift, we predict that the IF shift would also be reversed when the incident beam is in mixed linearly polarized state. The physical interpretation of these interesting phenomena lies in the reversed transverse angular momenta and reversed linear momenta. Although the spatial shifts have no direct relations with the $z$-component angular momentum ($z$-component angular momentum is not reversed in LHM), they are actually the outcomes of spin-orbit and orbit-orbit conversion. Therefore, the unreversed spatial shifts are indirect evidence of unreversed angular momentum of LHM. As for angular shifts, they are reversed at air-LHM interface and can be amplified by enhancing the vortex charge. This is direct evidence on the reversed linear momentum of LHM. Besides these qualitative
analysis, we also offer concrete expressions of the transverse linear and angular momenta, which explicitly reveal the physical picture of spin-orbit and orbit-orbit conversions. These momentum conversions are governed by $z$-component angular momentum conservation law.
\appendix*
\section{ELECTRIC FIELDS IN POSITION SPACE}\label{appendix}
In this appendix, we will give the full analytical expressions of the electric fields of reflected and transmitted beams. The electric fields in position space are given by inverse Fourier transformation
\begin{eqnarray}\label{Fouriertransformation}
u(x,y,z)&=&\frac{1}{2\pi}\int
dk_{x}dk_{y}\tilde{u}(k_x,k_y)\exp[i(k_xx+k_yy)]\nonumber\\
&&\times\exp\left[inkz-\frac{iz}{2nk}(k_x^2+k_y^2)\right].
\end{eqnarray}
We first apply Eq.~(\ref{Fouriertransformation}) to calculate the reflected fields. Substituting Eq.~(\ref{reflectedbeam}) into Eq.~(\ref{Fouriertransformation}), we get the reflected beams
\begin{eqnarray}
\mathbf{E}_r&=&\left\{\alpha
r_p+\frac{i}{k}\left[\alpha\frac{\partial r_p}{\partial
\theta_i}\frac{\partial}{\partial x_r}-\beta\left(r_s+r_p\right)\cot\theta_i\frac{\partial}{\partial y_r}\right]\right\}u_r\hat{\mathbf{x}}_r\nonumber\\
&+&\left\{\beta r_s+\frac{i}{k}\left[\beta\frac{\partial
r_s}{\partial\theta_i}\frac{\partial}{\partial x_r}+\alpha\left(r_s+r_p\right)\cot\theta_i\frac{\partial}{\partial y_r}\right]\right\}u_r\hat{\mathbf{y}}_r\nonumber\\
&+&\frac{i}{k}\left(\alpha r_p\frac{\partial}{\partial x_r}+\beta
r_s\frac{\partial}{\partial y_r}\right)u_r\hat{\mathbf{z}}_r,\label{reflectedbeam_position_space_1}\\
u_r&=&\frac{C_lkw_0}{2(z_R+iz_r)}\left[\frac{kw_0}{\sqrt{2}}\frac{x_r-i~\text{sgn}[l]y_r}{z_R+iz_r}\right]^{|l|}\nonumber\\
&&\times\exp\left[-\frac{k(x_r^2+y_r^2)}{2(z_R+iz_r)}+ikz_r\right].\label{reflectedbeam_position_space_2}
\end{eqnarray}
Note that the orbital angular momentum of reflected beam is $-l\hbar$ per photon. Eqs.~(\ref{reflectedbeam_position_space_1}) and (\ref{reflectedbeam_position_space_2}) are the full expressions of reflected electric fields.

We proceed to calculate the transmitted electric fields. Substituting Eq.~(\ref{transmittedbeam}) into Eq.~(\ref{Fouriertransformation}), we get the transmitted fields
\begin{eqnarray}
\mathbf{E}_t&=&\left\{\alpha
t_p-\frac{i}{k}\left[\alpha\eta\frac{\partial t_p}{\partial
\theta_i}\frac{\partial}{\partial x_t}+\beta\left(t_p-\eta t_s\right)\cot\theta_i\frac{\partial}{\partial y_t}\right]\right\}u_t\hat{\mathbf{x}}_t\nonumber\\
&+&\left\{\beta t_s-\frac{i}{k}\left[\beta\eta\frac{\partial
t_s}{\partial\theta_i}\frac{\partial}{\partial x_t}+\alpha\left(\eta t_p-t_s\right)\cot\theta_i\frac{\partial}{\partial y_t}\right]\right\}u_t\hat{\mathbf{y}}_t\nonumber\\
&+&\frac{i}{nk}\left(\alpha t_p\frac{\partial}{\partial x_t}+\beta
t_s\frac{\partial}{\partial
y_t}\right)u_t\hat{\mathbf{z}}_t.\label{transmittedbeam_position_space_1}
\end{eqnarray}
The formula of $u_t$ is rather lengthy. If $|l|$ is an even number, then the formula  is
\begin{widetext}
\begin{eqnarray}
u_t&=&\frac{C_l|n|kw_0}{2\pi}\frac{1}{\sqrt{(z_{R_x}+iz_t)(z_{R_y}+iz_t)}}\left[\frac{w_0}{\sqrt{2}}\right]^{|l|}\exp\left[-\frac{nk}{2}\left(\frac{x_t^2}{z_{R_x}+iz_t}+\frac{y_t^2}{z_{R_y}+iz_t}\right)+inkz_t\right]\nonumber\\
&&\times\bigg\{\sum_{m=0,2,4\cdots}^{|l|}C_{|l|}^{m}\left[\frac{-i\eta}{\sqrt{a}}\right]^{|l|-m}\left[\frac{\text{sgn}[l]}{\sqrt{c}}\right]^{m}\exp[-ab^2-cd^2]\Gamma\left[\frac{1+|l|-m}{2}\right]\Gamma\left[\frac{1+m}{2}\right]\nonumber\\
&&~~~\times~\leftidx{_1}F_1\left[\frac{1+|l|-m}{2},\frac{1}{2},ab^2\right]\leftidx{_1}F_1\left[\frac{1+m}{2},\frac{1}{2},cd^2\right]\nonumber\\
&&~~~+\sum_{m=1,3,5\cdots}^{|l|-1}C_{|l|}^{m}\left[\frac{-i\eta}{\sqrt{a}}\right]^{|l|-m}\left[\frac{\text{sgn}[l]}{\sqrt{c}}\right]^{m}4\sqrt{ac}bd\exp[-ab^2-cd^2]\Gamma\left[1+\frac{|l|-m}{2}\right]\Gamma\left[1+\frac{m}{2}\right]\nonumber\\
&&~~~\times~\leftidx{_1}F_1\left[1+\frac{|l|-m}{2},\frac{3}{2},ab^2\right]\leftidx{_1}F_1\left[1+\frac{m}{2},\frac{3}{2},cd^2\right]\bigg\};\label{transmittedbeam_position_space_2}
\end{eqnarray}
if $|l|$ is an odd number, then the formula  is
\begin{eqnarray}
u_t&=&\frac{C_l|n|kw_0}{2\pi}\frac{1}{\sqrt{(z_{R_x}+iz_t)(z_{R_y}+iz_t)}}\left[\frac{w_0}{\sqrt{2}}\right]^{|l|}\exp\left[-\frac{nk}{2}\left(\frac{x_t^2}{z_{R_x}+iz_t}+\frac{y_t^2}{z_{R_y}+iz_t}\right)+inkz_t\right]\nonumber\\
&&\times\bigg\{\sum_{m=0,2,4\cdots}^{|l|-1}C_{|l|}^{m}\left[\frac{-i\eta}{\sqrt{a}}\right]^{|l|-m}\left[\frac{\text{sgn}[l]}{\sqrt{c}}\right]^{m}2\sqrt{a}b\exp[-ab^2-cd^2]\Gamma\left[1+\frac{|l|-m}{2}\right]\Gamma\left[\frac{1+m}{2}\right]\nonumber\\
&&~~~\times~\leftidx{_1}F_1\left[1+\frac{|l|-m}{2},\frac{3}{2},ab^2\right]\leftidx{_1}F_1\left[\frac{1+m}{2},\frac{1}{2},cd^2\right]\nonumber\\
&&~~~+\sum_{m=1,3,5\cdots}^{|l|}C_{|l|}^{m}\left[\frac{-i\eta}{\sqrt{a}}\right]^{|l|-m}\left[\frac{\text{sgn}[l]}{\sqrt{c}}\right]^{m}2\sqrt{c}d\exp[-ab^2-cd^2]\Gamma\left[\frac{1+|l|-m}{2}\right]\Gamma\left[1+\frac{m}{2}\right]\nonumber\\
&&~~~\times~\leftidx{_1}F_1\left[\frac{1+|l|-m}{2},\frac{1}{2},ab^2\right]\leftidx{_1}F_1\left[1+\frac{m}{2},\frac{3}{2},cd^2\right]\bigg\},\label{transmittedbeam_position_space_3}
\end{eqnarray}
\end{widetext}
where, $C_{|l|}^m$ is the binomial coefficient, $\Gamma$ is the Gamma function, $\leftidx{_1}F_1$ is the Kummer confluent hypergeometric function, $a=\eta^2w_0^2/4+iz_t/2nk$, $b=ix_t/2a$, $c=w_0^2/4+iz_t/2nk$, $d=iy_t/2c$. Equations~(\ref{transmittedbeam_position_space_1})-(\ref{transmittedbeam_position_space_3}) fully describe the transmitted vortex beams. Although
Eqs.~(\ref{transmittedbeam_position_space_2}) and (\ref{transmittedbeam_position_space_3}) are cumbersome, we still manage to get the electric fields for several low order vortex beams
\begin{eqnarray}
u_t^{l=1}&\propto&\left(\frac{\eta
x_t}{z_{R_x}+iz_t}+\frac{iy_t}{z_{R_y}+iz_t}\right),\label{vortex_1}\\
u_t^{l=-1}&\propto&\left(\frac{\eta
x_t}{z_{R_x}+iz_t}-\frac{iy_t}{z_{R_y}+iz_t}\right),\label{vortex_-1}\\
u_t^{l=2}&\propto&\bigg[\left(\frac{\eta
x_t}{z_{R_x}+iz_t}+\frac{iy_t}{z_{R_y}+iz_t}\right)^2\nonumber\\
&&+\frac{iz_t(1-\eta^2)}{nk(z_{R_x}+iz_t)(z_{R_y}+iz_t)}\bigg],\\
u_t^{l=-2}&\propto&\bigg[\left(\frac{\eta
x_t}{z_{R_x}+iz_t}-\frac{iy_t}{z_{R_y}+iz_t}\right)^2\nonumber\\
&&+\frac{iz_t(1-\eta^2)}{nk(z_{R_x}+iz_t)(z_{R_y}+iz_t)}\bigg].
\end{eqnarray}
\begin{acknowledgments}
This work is supported by the projects of the National Natural
Science Foundation of China (Grant No. 61025024 and 11074068).
\end{acknowledgments}

\end{document}